\documentclass{aa}
\usepackage{graphicx}
\usepackage{txfonts}
\usepackage{booktabs,caption}
\usepackage{graphicx}
\usepackage{multicol}
\DeclareMathOperator{\sech}{sech}
\begin{document}

   \title{The constraining effect of gas and the dark matter halo on the vertical stellar distribution of the Milky Way}
        \titlerunning {The constraining effect of gas and the dark matter halo}

   \author{S. Sarkar
          \inst{1}
                %\email{suchira@iisc.ac.in}\\
          \and
           C.J. Jog
                \inst{1}
                %\email{cjjog@iisc.ac.in}
           }

   \institute { Department of Physics,
    Indian Institute of Science, Bangalore 560012, India \\
              \email{suchira@iisc.ac.in}\\
            \email{cjjog@iisc.ac.in}
              }

\abstract
{We study the vertical stellar distribution of the Milky Way thin disk in detail with particular focus on the outer disk. We treat the galactic disk as a gravitationally coupled, three-component system consisting of stars, atomic hydrogen gas, and molecular hydrogen gas in the gravitational field of the dark matter halo. The self-consistent vertical distribution for stars and gas in such a realistic system is obtained for radii between 4-22 kpc. 
 The inclusion of an additional gravitating component constrains the vertical stellar distribution toward the mid-plane,
so that the mid-plane density is higher, the disk thickness is reduced, and the vertical density profile is steeper than in the one-component, isothermal, stars-alone case.  
 We show that the stellar distribution is constrained mainly by the gravitational field of gas and dark matter halo in the inner and the outer Galaxy, respectively.
We find that the thickness of the stellar disk 
(measured as the HWHM of the vertical density distribution) increases with radius, flaring
steeply beyond R=17 kpc. The disk thickness is reduced by a factor of 3-4  in the outer 
Galaxy as a result of the gravitational field of the halo, which may help the disk resist 
distortion at large radii. The disk would flare even more if the effect of  dark matter halo were not taken into account. Thus it is crucially important to include the effect of 
the dark matter halo  when determining the vertical structure and dynamics of a galactic 
disk in the outer region.}


   \maketitle

\section{Introduction}
The stars in a spiral galaxy are distributed in a thin disk with a finite vertical extent. An 
important property that characterizes a gravitating disk is its vertical density profile, $\rho(z),$ where $\rho$ is the mass density and $z$ is
along the direction normal to the disk plane. This is because the mass distribution and dynamics in a gravitating system are 
inter-related (e.g., Binney \& Tremaine 1987). For a one-component disk, the self-consistent vertical density distribution for an isothermal disk is given as a $sech^2 (z)$ distribution (Spitzer 1942).

A real galaxy disk, however, consists of stars and interstellar gas. The gas contains  10-15 \% of the disk mass (Young \& Scoville 1991, Binney \& Merrifield 1998).
However, because of its lower dispersion, the gas forms a thin layer and hence can affect the vertical distribution of stars significantly despite its low mass faction. The self-consistent vertical structure of such a multi-component disk (consisting of stars, atomic hydrogen gas, $\mathrm{HI,}$ and molecular hydrogen gas, $\mathrm{H_2}$) where the components are gravitationally coupled was studied by Narayan \& Jog (2002b) for the Galaxy for radii $< 12$ kpc, where it was shown that the vertical distribution of each component is affected by the others. 
This approach has subsequently been used by many papers in the literature, including those that  derive the disk properties from observations (e.g., Comeron et al. 2011) as it gives a more accurate result 
for the disk thickness. 

In this paper, we adopt the method proposed by Narayan \& Jog (2002b) and extend their study to the outer disk of the Milky Way between radii of 12-22 kpc. The aim is to study the dynamical effect of gas as well as of the dark matter halo on the vertical stellar profile of the thin disk. The density of the disk components as well as the dark matter halo density decrease with radius, but in the outer parts of a galaxy,
the halo is likely to dominate the dynamics. Further motivation for our study comes from recent observations that show that the stellar disk flares in the outer Galaxy (e.g., Momany et al. 2006, Lopez-Corredoira \& Molgo 2014).
We show that the gravitational effect of any coupled component causes the stellar distribution to be constrained  toward the mid-plane, thus reducing the disk thickness and making the density profile steeper than the single-component stars-alone case. We find that 
the constraining effect of  dark matter halo starts to dominate beyond 14 kpc. The thickness of the stellar distribution increases with radius, flaring steeply beyond 17 kpc. If the effect of dark matter halo were not included, the stellar disk thickness would flare even more. Thus the dark matter halo plays a pivotal role in confining the stellar distribution closer to the mid-plane in the outer Galactic disk.

Section 2 contains the formulation and numerical solution of equations, and the input parameters used. Section 3 contains results for the constraining effect of gas and dark matter halo. Section 4 contains 
a brief discussion to show that the gas distribution is also vertically confined by the effect of stars and the halo. The main conclusions are given in Section 5.
\section{Formulation of the equations and numerical solution}
For a one-component disk, the thickness is determined by the balance between its self-gravity and vertical pressure (Spitzer 1942).
To study the gravitationally coupled disk components (stars, HI and H$_2$ gas) in the field of dark matter halo,  we follow the treatment 
as given in Narayan \& Jog (2002b) and solve the following coupled equations to obtain the net density distribution:
\begin{equation}
\begin{split}
\frac{\mathrm{d}^{2}\rho_{i}}{\mathrm{d}z^{2}} & = \frac{\rho_{i}}{\left\langle(V_{z})^{2}_{i}\right\rangle}\left[-4\pi G\left(\rho_{\mathrm{s}}+\rho_{\mathrm{HI}}+\rho_{\mathrm{H_{2}}}\right)+\frac{\mathrm{d}(K_{z})_{\mathrm{DM}}}{\mathrm{d}z}\right]  \\
                           &   +\frac{1}{\rho_{i}}\left(\frac{\mathrm{d}\rho_{i}}{\mathrm{d}z}\right)^{2}, \label{eq:1}
\end{split}
,\end{equation}
\noindent where $i = 1,2,$ and $3$. Here $v_z$ is the velocity and $|K_z|$ is the force per unit mass, both along the $z$-axis. The galactocentric cylindrical coordinates ($R, z, \phi$) are used, and $z=0$ denotes the galactic mid-plane. We assume each component to be isothermal
so that the vertical velocity dispersion $(\sigma_z)_i$ (=$\langle(v_z)_i^2\rangle^{1/2}$)  is taken to be constant along the $z$ axis.
This equation was solved with a fourth-order Runge-Kutta method in an iterative way for each component with fifth decimal convergence, following the same approach as in (Narayan \& Jog 2002b).  

The formulation described above is general. We next apply it to the Milky Way since various parameters such as the surface density and the velocity dispersion are known for it observationally.
For the dark matter halo we consider a pseudo-isothermal profile with the parameters ($R_c= 5$ kpc, and $V_c = 220 $ km s$^{-1}$) taken as in Narayan \& Jog (2002b), and the derived potential being
\begin{equation}
\psi_{\mathrm{DM}}(r)=-V_{\mathrm{c}}^{2}\left[1-\frac{1}{2}\log\left(R^{2}_{\mathrm{c}}+r^{2}\right)-\frac{R_{\mathrm{c}}}{r}\tan^{-1}\left(\frac{r}{R_{\mathrm{c}}}\right)\right]. \label{eq:2}
\end{equation}
\noindent where $r$ is the radius in spherical coordinates. Here $R_c$ is the core radius and $V_c$ is 
the limiting rotation velocity.        Next, using cylindrical coordinates, we write $\mathrm{d}(K_{z})_{\mathrm{DM}}/\mathrm{d}z =-\partial^{2}\psi_{\mathrm{DM}}/\partial z^{2} $ as the halo contribution in Eq.1 (as given in Narayan \& Jog (2002b) with a negative sign multiplying the total expression, since the negative sign was missed in the expression for the potential in that paper).
  
The stellar surface density values were taken from the Galaxy mass model of Mera et al. (1998).        
The radial stellar velocity dispersion values were taken from the observational values obtained for the Galaxy by Lewis \& Freeman (1989) up to 16 kpc. This follows an exponential curve:
\begin{equation}
\sigma_{R}= 105 \exp(-R/8.7\mathrm{kpc}) ~\mathrm{km~s^{-1}} \label{eq:3}
.\end{equation}
From these the corresponding vertical velocity dispersion values $\sigma_{z}$ were calculated by assuming the vertical-to-planar ratio to be 0.45 
(Dehnen \& Binney 1998; Mignard 2000).
Beyond 17 kpc, the $\sigma_{z}$  value was kept constant equal to its last value of 7.5~$\mathrm{km~s^{-1}}$ so as to be higher than the $\sigma_{z}$ value of HI. The physical reason for this assumption is given next. 
We note that stars form from the collapse of gas clouds, therefore the stellar velocity dispersion $\sigma_{z}$ 
cannot be lower than that of the dispersion within the gas. 
 Once the stars form, they gain random kinetic energy 
 via gravitational scattering off clouds or spiral structure in the disk, hence the velocity dispersion shows a systematic increase with stellar age in the standard scenario of stellar heating (e.g., Binney \& Tremaine 1987). 

For HI gas between R = 4 to 14 kpc, the observed surface density values from Scoville \& Sanders (1987) were used.
Beyond that radius, the values were taken from Levine et al. (2006), who studied the outer Galactic HI disk and measured the disk surface density,$\Sigma_{\mathrm{HI}}$, which they found was best  fit by the following exponential curve:
\begin{equation}
\Sigma_{\mathrm{HI}}\left(R\right)=4.5\exp[-\left(R-14\mathrm{kpc}\right)/4.3\mathrm{kpc}]~M_{\odot}\mathrm{pc^{-2}}. \label{eq:4}
\end{equation}
The velocity dispersion $\sigma_{z}$  for HI gas is measured only in the inner Galactic disk, see Narayan et al. (2005) for a detailed discussion,  the salient points from which are mentioned here. From R= 4 to 12 kpc, the value of $\sigma_{z}$ for HI  is taken  as 8~$\mathrm{km~s^{-1}}$ based on the values given by Spitzer (1978) for the Galaxy and by Lewis (1984) for nearly 200 face-on galaxies. The velocity is observed to decrease and then saturate to around 7 $\pm$ 1 km s$^{-1}$ in the outer parts (e.g., Kamphuis 1993; Dickey 1996).
Beyond R=12 kpc,  we therefore consider the velocity to taper off with a slope of $-$0.2~$\mathrm{km~s^{-1}~kpc^{-1}}$ up to 17 kpc and then  assume it to be constant at the last value of 7~$\mathrm{km~s^{-1}}$ for outer radii beyond that. 
 Since the observed values of $\sigma_{z}$ for HI and stars in the outer disk are not very well-determined, we do not go beyond R=22 kpc in this analysis.
For $\mathrm{H_{2}}$ gas, the observed surface density and velocity dispersion values are taken from Scoville \& Sanders (1987). 
\begin{table}
\caption{Surface density of stars and gas vs. radius in the Milky Way}
\label{table:1}
\centering
  \begin{tabular}{l l l l}
\hline \hline
Radius & $\Sigma_{\mathrm{Stars}}$\tablefootmark{a} & $\Sigma_{\mathrm{HI}}$\tablefootmark{b} & $\Sigma_{\mathrm{H_{2}}}$\tablefootmark{b} \\
(kpc) & $(M_{\sun}\mathrm{pc^{-2}})$ & $(M_{\sun}\mathrm{pc^{-2}})$ & $(M_{\sun}\mathrm{pc^{-2}})$ \\
\hline
4.0 & 183.6 & 4.6 & 13.1\\
6.0 & 98.3 & 4.6 & 10.8\\
8.5 & 45.0 & 5.5 & 2.1\\
10.0 & 28.2 & 5.5 & 0.8\\
12.0 & 15.1 & 5.5 & 0.4\\
14.0 & 8.1 & 4.0 & $\cdot\cdot\cdot$\\
16.0 & 4.3 & 2.8 & $\cdot\cdot\cdot$\\
18.0 & 2.3 & 1.8 & $\cdot\cdot\cdot$\\
20.0 & 1.2 & 1.1 & $\cdot\cdot\cdot$\\
22.0 & 0.66 & 0.70 & $\cdot\cdot\cdot$\\
\hline
\end{tabular}
\tablefoot{
\tablefoottext{a}{Mera et al. 1998}\\
\tablefoottext{b}{Scoville \& Sanders 1987}}
   \end{table}

\section{Results}
\subsection{Vertical density distribution of stars}
\begin{figure*}
\centering
\includegraphics[height=2.0in,width=2.70in]{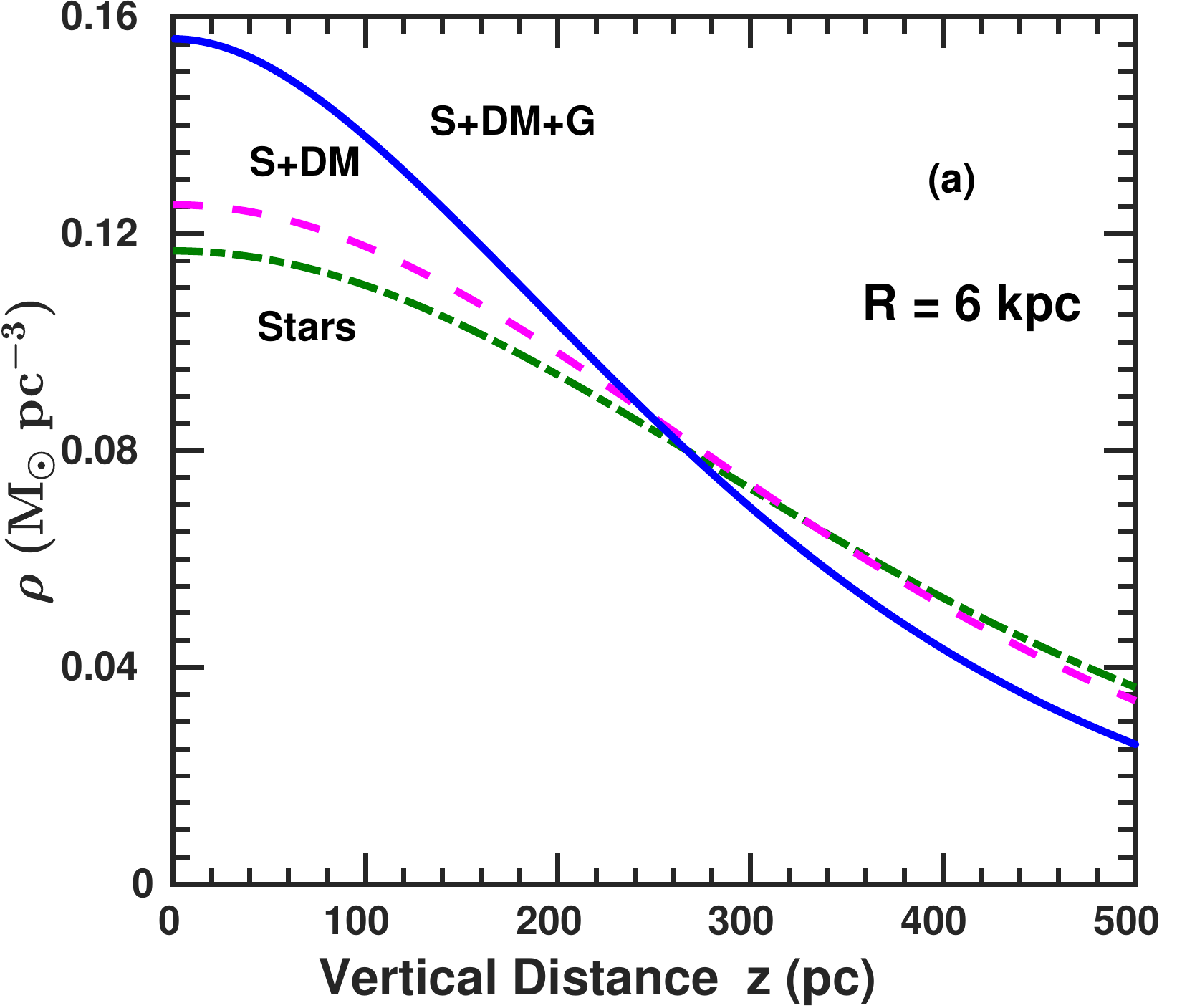}
\medskip
\includegraphics[height=1.98in,width=2.65in]{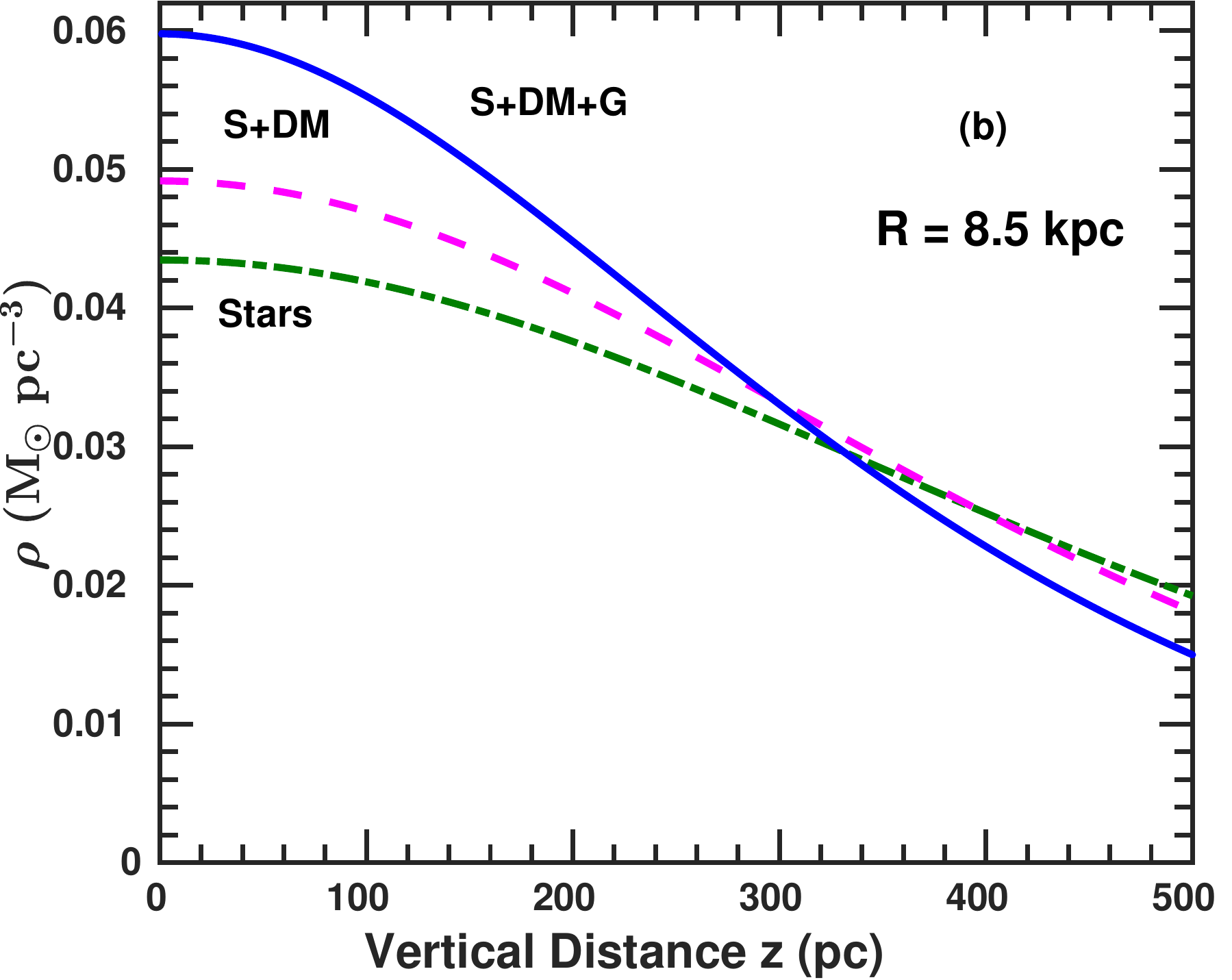}
\bigskip
\includegraphics[height=2.0in,width=2.65in]{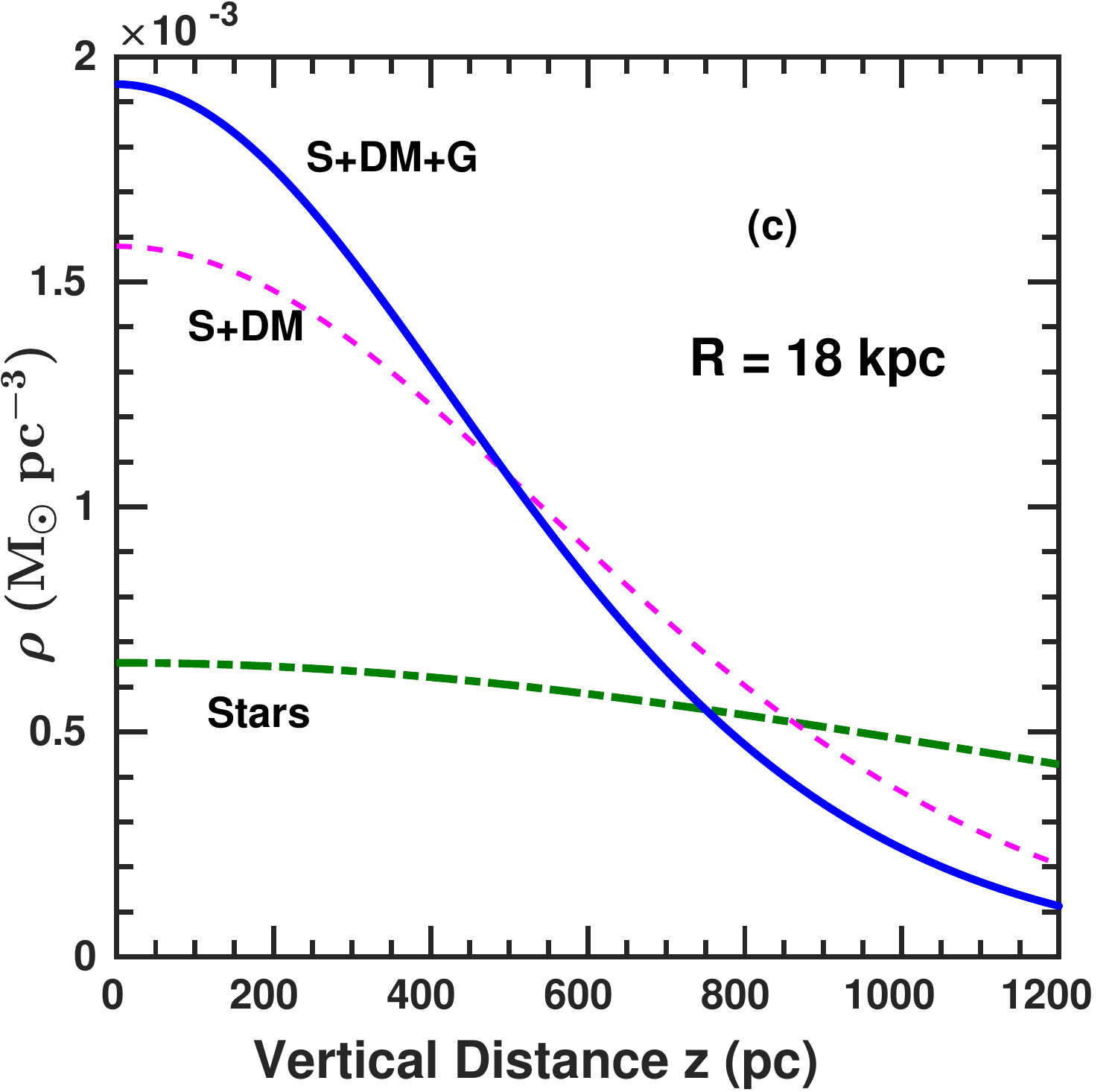}
\medskip
\includegraphics[height=2.0in,width=2.55in]{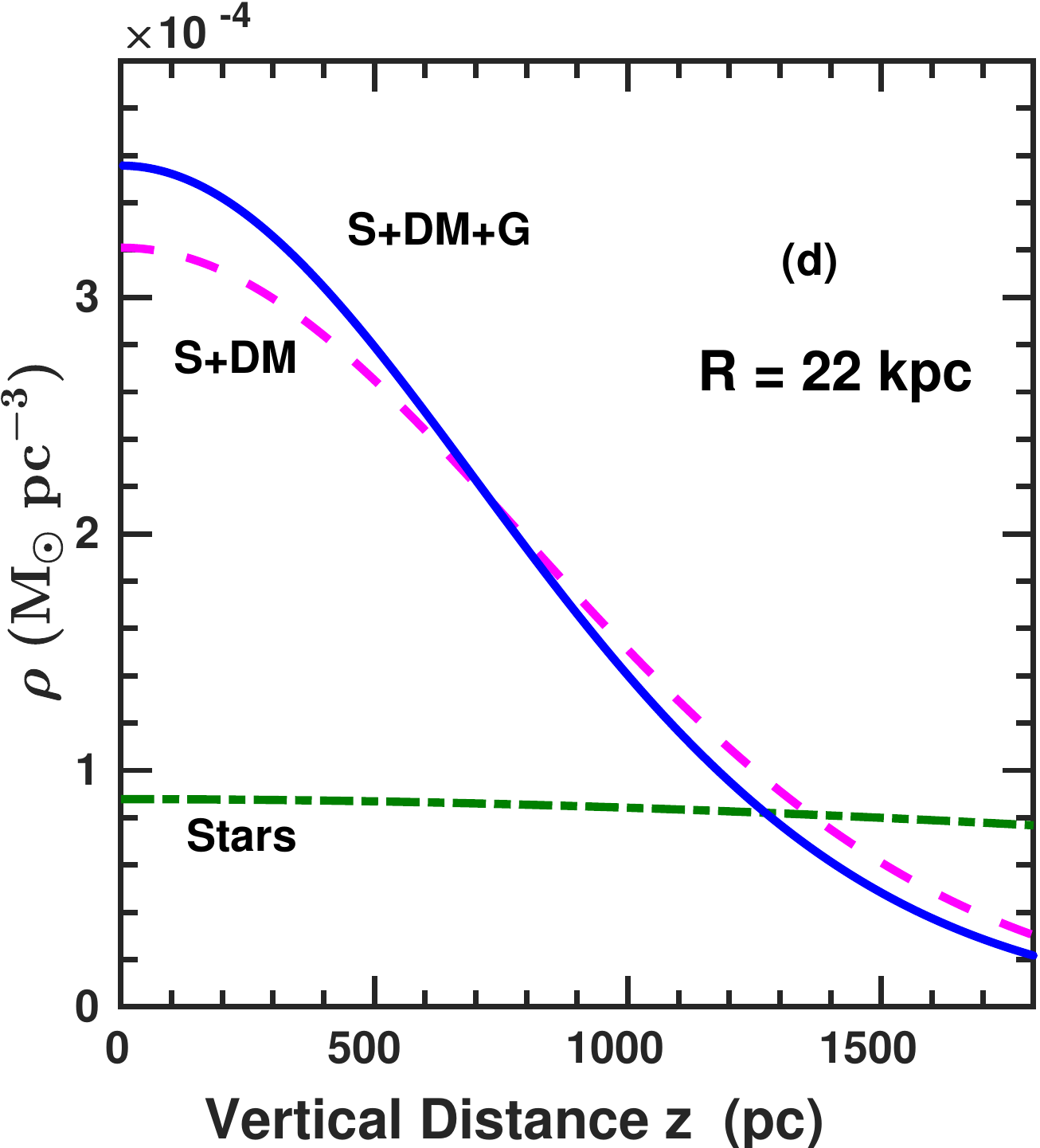}
\bigskip
\caption{Vertical density distribution of stars vs. $z$ at four different radii, $R=$ 6 and 8.5 kpc in the inner Galaxy (a,b) and at $R=$ 18 and 22 kpc in the outer Galaxy (c,d). The three curves represent the density distribution of stars in the gravitational field of stars-alone, stars plus dark matter halo, and stars plus dark matter halo plus gas cases. The addition of other gravitating components (gas and halo) increases the mid-plane stellar density value that causes the scale height to become smaller and the curve to become steeper. The constraining effect is mainly due to gas in the inner Galaxy (a,b); while in the outer Galaxy (c,d), the dark matter halo has the dominant constraining effect. }
\label{label1}
\end{figure*}

The numerical  solution of the coupled equation as given by Eq.(1) gives the self-consistent vertical density distribution at a given radius. To study its nature in detail from the inner to the outer Galaxy, we performed the calculation from $R=$4-22 kpc at an interval of 2 kpc. Instead of 8 kpc, we considered 8.5 kpc as it denotes the region of the solar neighborhood. The inner Galaxy limit was chosen to be 4 kpc because below this radius the bulge and the bar contribution may dominate the vertical luminosity. 
The resulting stellar profiles are shown in Fig. 1 for four selected radii, $R=$ 6, 8.5, 18, and 22 kpc. This choice will help  bring out the dynamical result that different components (gas or the dark matter halo) dominate at different radii, as discussed below.

In the inner part of the Galaxy, for instance, at $R$ =6 or 8.5 kpc, the surface density of stars is high and its self-gravity dominates  the gravity of gas and the halo. Interestingly, even though the gas contains only $\sim 10-15 \% $ of mass in the disk, it has a significant constraining effect on the net stellar density profile  as can be seen from the increased stellar mid-plane density and reduced stellar thickness, see the three distributions in Fig 1 (a,b).
In the outer Galaxy, for example, at R=18 kpc, the stellar surface density is low, and the stars-alone case therefore gives rise to an extended, diffuse vertical stellar distribution. When halo gravity is included, it strongly constrains the vertical stellar distribution,  compared to the effect of gas (Fig. 1 c,d).
Even farther out at R= 22 kpc, the effect of  dark matter halo is more prominent. Thus we conclude that with increasing radii, the dark matter halo is mainly responsible for shaping the stellar density distribution.  

Thus our results show that an additional component that can exert an  extra gravitational force increases the mid-plane density compared to the one-component case. For a constant surface density, it therefore also constrains the whole distribution to be closer  
 toward the plane, which thus has a smaller thickness.  The shape of the density profile is also altered and becomes steeper.

To give a physical explanation for the results shown in Fig. 1, we first calculated the vertical gravitational force per unit mass ,$|K_{z}|$, exerted by the stars, HI, $\mathrm{H_{2}}$ , and the dark matter halo, treated as independent components, as a function of  $z$. For a given disk component,  $K_{z}$ and the corresponding density distribution $\rho(z)$ can be obtained analytically by solving the force equation and the Poisson equation together to be (Spitzer 1942)
\begin{equation}
|K_{z}|=2\frac{\left\langle v^{2}_{z}\right\rangle}{z_{0}}\tanh\left(\frac{z}{z_{0}}\right)\:\:\: ; \: \: \:\rho(z) = \rho_{0} sech^{2}(z/z_{0})\label{eq:6} 
.\end{equation}
 Here $\rho_0$ is the mid-plane density and the constant $z_0$ is given by
\begin{equation}
z_0 = \left(\frac{{\left\langle v^{2}_{z}\right\rangle}}{2\pi G \rho_{0}}\right)^{1/2}\label{eq:7}
.\end{equation}

For each disk component, using the observed surface density at a given radius as the boundary condition, the density $\rho(z)$ (Eq. (5)) was integrated to obtain both $\rho_0$ and $z_0$ simultaneously. The force due to the dark matter halo in cylindrical coordinates was obtained from the expression for its potential (Eq. (2))  to be\begin{equation}
\begin{split}
\left(K_{z}\right)_{\mathrm{DM}} & = -\frac{V^{2}_{\mathrm{c}}z}{\left(R^{2}+R^{2}_{\mathrm{c}}+z^{2}\right)}+\frac{V^{2}_{\mathrm{c}}R_{\mathrm{c}}z}{\left(R^{2}+z^{2}\right)^{3/2}}\tan^{-1}\left(\frac{\sqrt{R^{2}+z^{2}}}{R_{c}}\right) \\
                 & -\frac{V^{2}_{\mathrm{c}}z}{\left(R^{2}+z^{2}\right)\left(1+\frac{R^{2}+z^{2}}{R^{2}_{\mathrm{c}}}\right)}.    \label{eq:8}
\end{split}
.\end{equation} 

In Fig.2  we plot $|K_{z}|$ for the various components at R= 6 kpc and R= 18 kpc.
At R=6 kpc, the gravitational force due to $\mathrm{H_2}$ gas is strong, $\sim 30 \%$ of that due to stars, for $|z| < 150$ pc. The force due to the dark matter halo is smaller, and we checked that it becomes more important than the force due to gas only beyond z $\sim 800$ pc. On the other hand, in the outer Galaxy, at R=18 kpc, the magnitude of the vertical force $|K_{z}|$ due to the halo dominates  that due to stars and HI gas from the lowest $z$ values.

\begin{figure*}
\centering
\includegraphics[height=2.0in,width=2.8in]{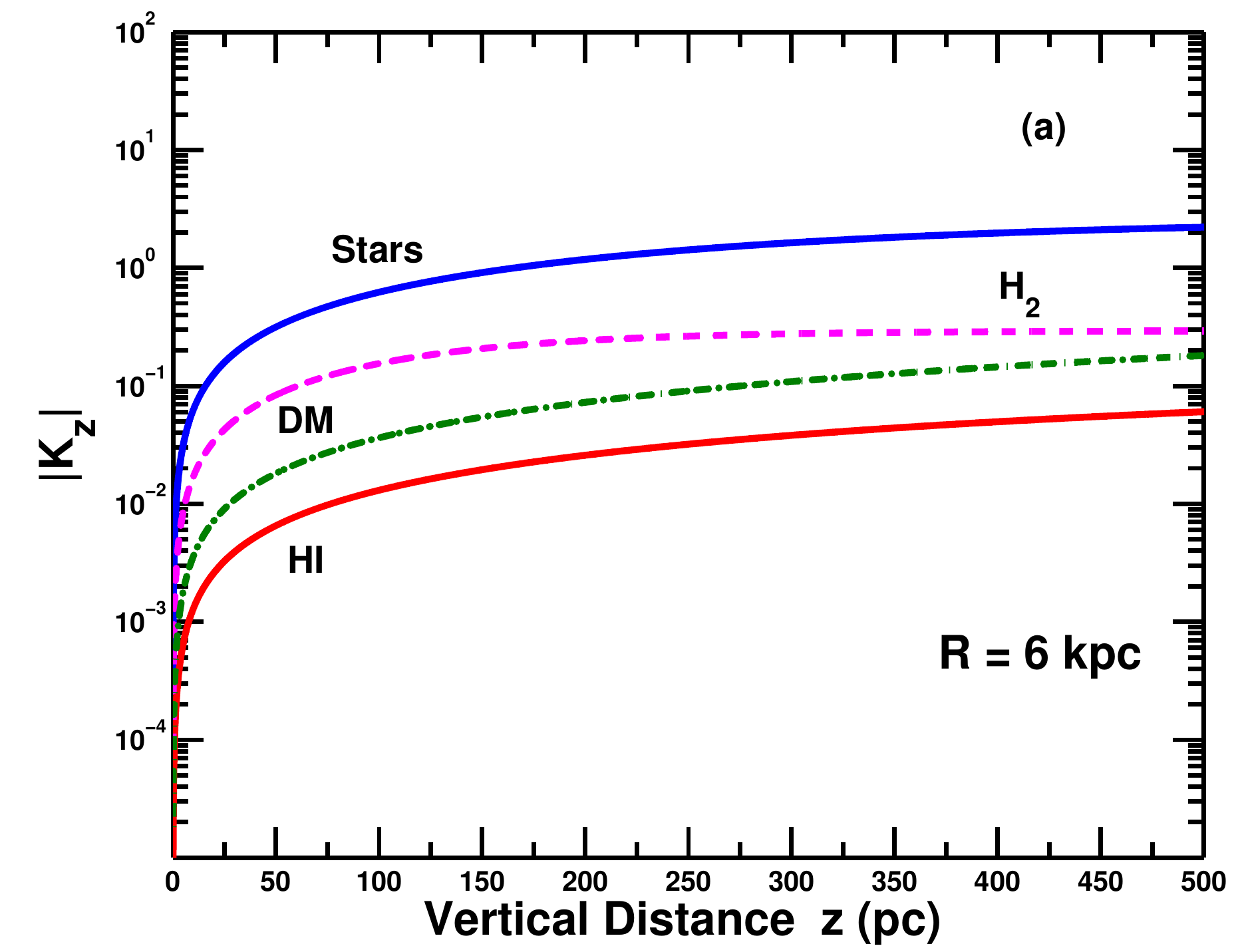}
\medskip
\includegraphics[height=2.05in,width=2.8in]{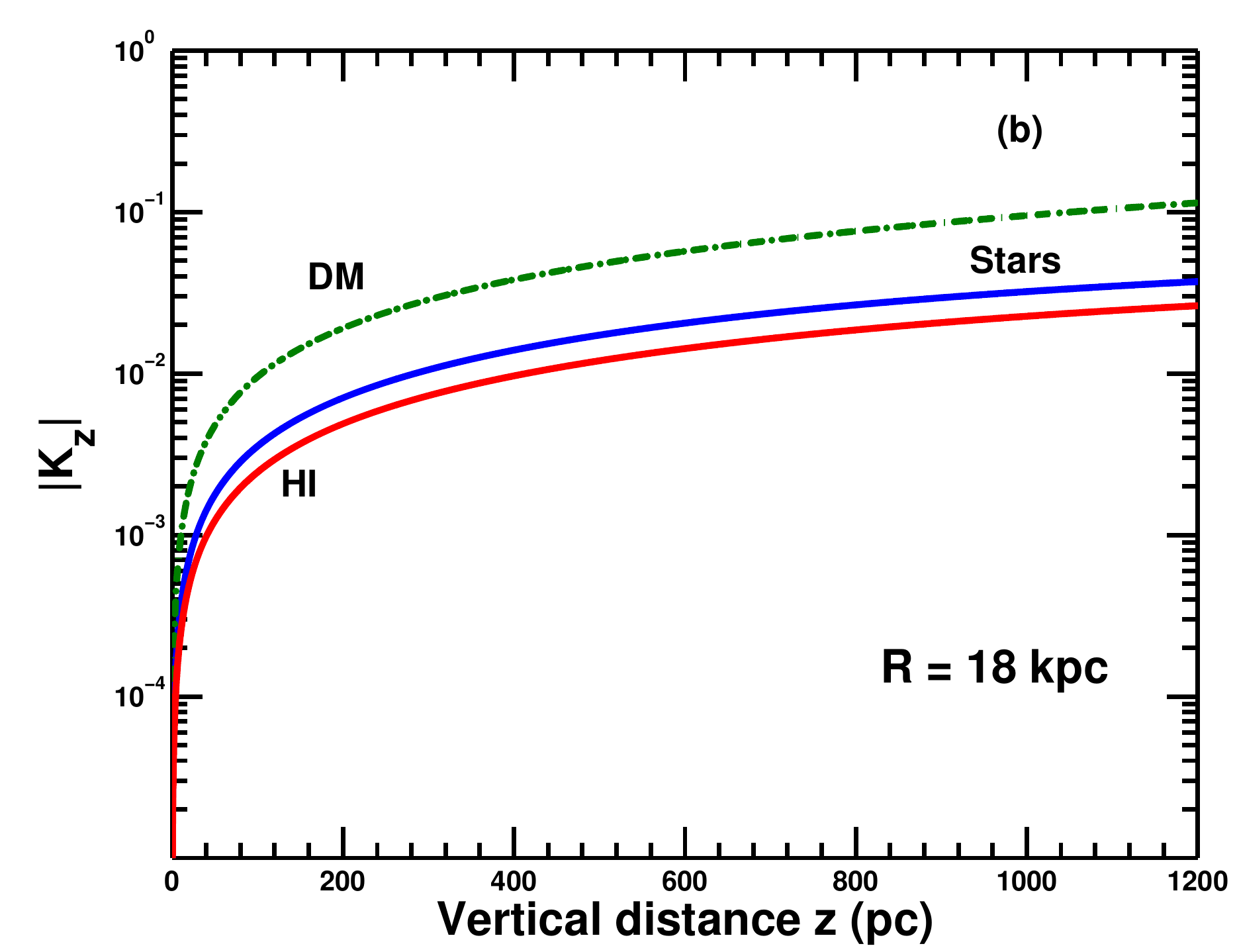}
\bigskip
\caption{Vertical gravitational force exerted by each component treated separately vs. $z$ at R = 6 kpc (a) and 18 kpc (b). 
At R=6 kpc, the stellar distribution is mainly affected by the gas gravity, while at R=18 kpc, the force due to the dark 
matter halo dominates and hence strongly affects the stellar distribution.}
\label{label2}
\end{figure*}

\begin{figure*}
\centering
\includegraphics[height=1.92in,width=2.8in]{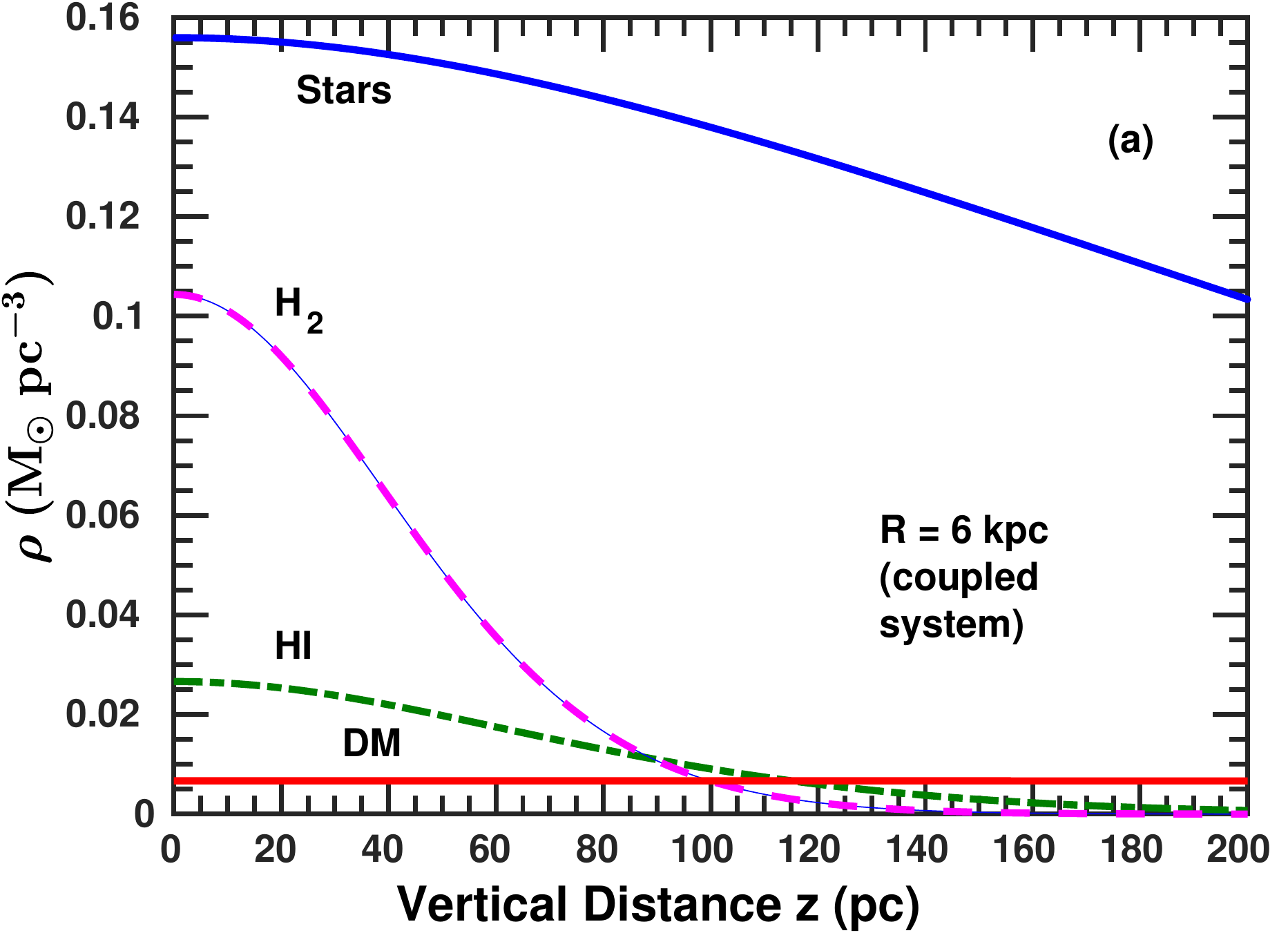}
\medskip
\includegraphics[height=2.0in,width=2.8in]{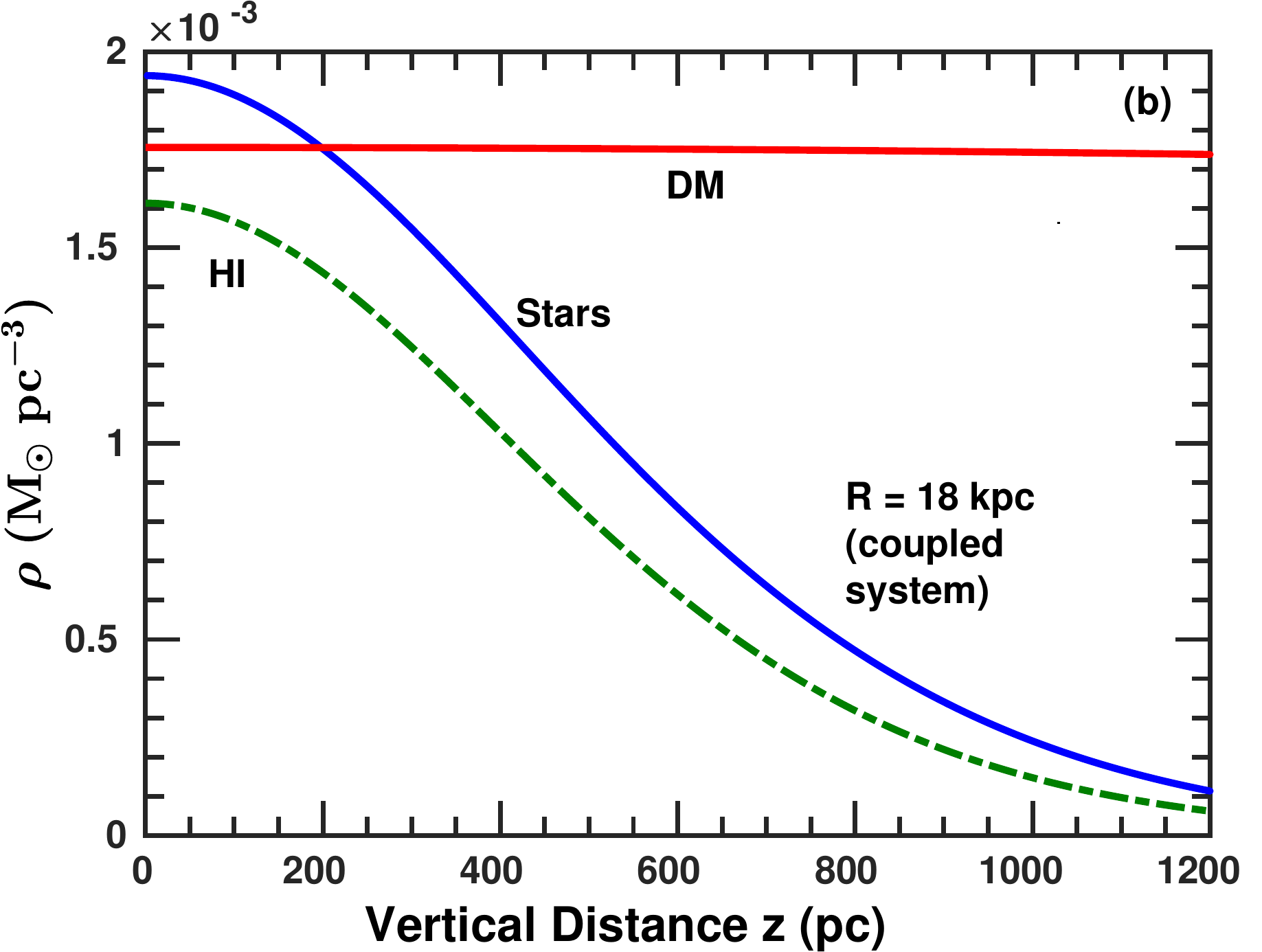}
\bigskip
\caption{Density for stars and gas for the coupled system and the corresponding term due to the dark matter halo $|(dK_z/dz)|/4 \pi G $ (Eq. 1) vs. $z$,
at R= 6 kpc (a) and at R=18 kpc (b). 
Because of its lower dispersion, the mid-plane $\mathrm{H_2}$ gas density is comparable to the stellar value, and hence despite its low surface density, it 
 can significantly affect the stellar distribution at R= 6 kpc. In the outer Galaxy, the halo gravitational effect dominates, hence it is mainly responsible for the determination of the vertical stellar distribution.}
\label{label3}
\end{figure*}

For the coupled system, the corresponding force components are not known, therefore we instead obtained the gradient of the force with $z$, that is,( $dK_{z}/dz$) in terms of the mass density 
$\rho(z)$ for each disk component (Banerjee \& Jog 2007).  For the halo we plot the value of $-(dK_{z}/dz)\times(1/4\pi G)$. The resulting values are plotted for R=6 kpc and R=18 kpc in Fig. 3.
Fig 3a confirms the strong effect of gas in constraining the vertical stellar distribution in the inner Galaxy. In an earlier study, a similar constraining effect on the disk around a molecular cloud complex  (of $\sim 10^7M_{\sun}$ mass and with an extent of $\sim 200$ pc)  was shown by  Jog \& Narayan (2001). 
 At R=6 kpc, the dark matter halo density is lower than the gas density for $|z| < 120 pc$ and lower than the stellar density until about 700 pc (not shown in the figure), and therefore it does not play an important role.
On the other hand, in the outer Galaxy, at R=18 kpc (Fig 3b),  
the halo dominates  the gas at all z values and  stars at z $>$ 200 pc. This shows that the halo has a very strong constraining effect  that increases progressively at larger radii. This effect is also evident in the higher increase in the mid-plane density at larger radii that occurs when  halo gravity is included (Fig. 1c,d).

\subsection{Measuring the constraining effect: Disk thickness}
\subsubsection{ Half-width at half-maximum}
\begin{figure}
\centering
\includegraphics[height=2.0in,width=2.6in]{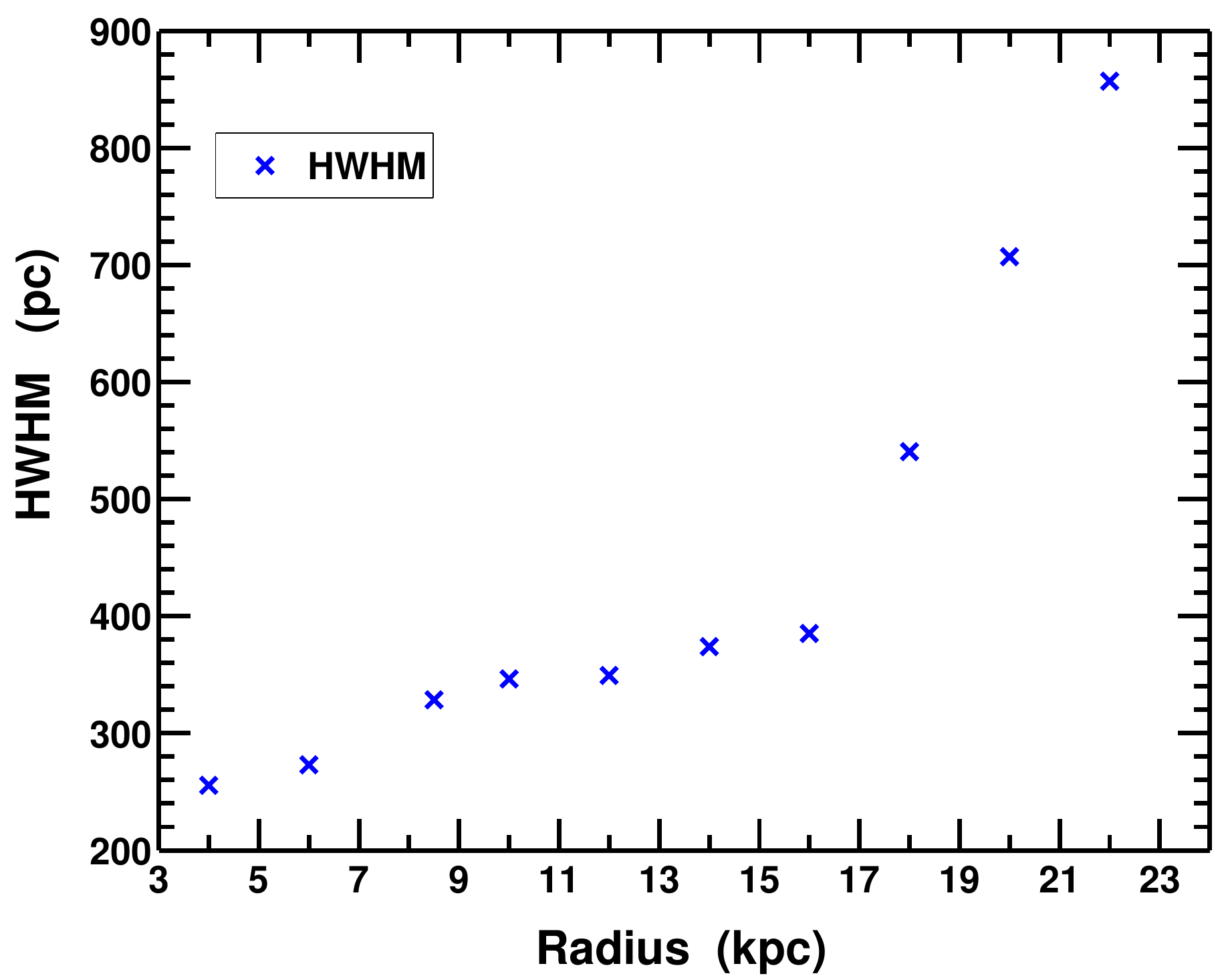}
\caption{Results for the variation in HWHM of the vertical stellar density distribution in the joint potential of the disk and dark matter halo vs. radius in the Galactic disk. The stellar disk thickness increases gradually until about 17 kpc and then flares beyond that in the outer disk.}
\label{label4}
\end{figure}
To estimate quantitatively the constraining effect of different components, we next calculate the half-width at half-maximum (HWHM) of the vertical stellar distribution, taken to be an indicator of disk thickness, see Table 2 and Fig. 4. This quantity is also sometimes referred to as the vertical scale height (e.g., Narayan \& Jog (2002 a,b)). 
In the inner Galaxy, for instance, at R=6 kpc,  the HWHM is reduced by $6\%$ when the halo is included compared to the stars-alone case, while it is reduced by $26\%$ in presence of both halo and gas. Thus the gravitational force due to gas is mainly responsible for decreasing the thickness.
On the other hand, in the outer Galaxy, for example, at R=18 kpc, including the halo has a very strong effect that leads to a  decrease in the stellar HWHM by a factor of $\sim 2.3, $ while including gas decreases it by another $\sim 20 \%$ only.
The stellar disk by itself would flare by a factor of 14 from R=4 to 22 kpc.  Thus the inclusion of halo mainly limits the disk to a net thickness of $< 1$ kpc even at R=22 kpc. This would help the disk resist distortion due to external perturbations.

In a previous study, the dense, compact halo in low surface density galaxies such as UGC 7321 
was shown to be mainly responsible for reducing the disk thickness (Banerjee \& Jog 2013).  
Here we have shown  that in general, when the halo term is larger than the stellar term in the r.h.s. of Eq. (1),
 the halo will have a dominant constraining effect as in the outer Milky Way disk.
\begin{table}
\caption{HWHM of vertical stellar density distribution vs. radius}
\label{table:2}
\centering
\begin{tabular}{l l l l}
\hline \hline
Radius & Stars-alone & Stars+Halo & Stars+Halo+Gas \\
(kpc) & (pc) & (pc) & (pc) \\ \hline
4.0  & 313.6  & 301.6  & 255.6 \\
6.0  & 370.8  & 347.7  & 273.2\\
8.5  & 456.7  & 407.6  & 328.7 \\
10.0 & 515.4  & 440.4  & 346.5\\
12.0 & 606.2  & 478.1  & 349.5\\
14.0 & 708.3  & 501.7  & 374.1\\
16.0 & 822.2 & 507.9  & 385.4\\
18.0 & 1568.9 & 672.9  & 540.6\\
20.0 & 2721.8 & 824.1  & 703.7\\
22.0 & 4279.9 & 957.9  & 857.2\\
\hline
\end{tabular}
\end{table}

Thus, we have shown that the detailed balance given by Eq. (1) in a multi-component disk results in  a moderate increase in disk thickness of 40$\%$ up to R= 12 kpc, and the disk flares steeply beyond R=17 kpc.
Since the dispersion does not fall rapidly at radii beyond R=17 kpc (by assumption based on the physical argument, see Section 2),  this results in a rapid flaring beyond R=17 kpc.
The stellar flaring obtained could be even higher if a higher gas dispersion value of 10 ~$\mathrm{km~s^{-1}}$ as in Tamburro et al (2009) 
is adopted.
In our model, the coupled equations are such (see Eq. (1)) that the choice of the velocity dispersion, $\sigma_{z}$,  mainly  affects the density profile for that particular component, and has only a small indirect effect on the other components. However, because we have set the physical condition that the stellar velocity dispersion value should be greater than for gas (see Section 2),   a higher gas dispersion will also imply a higher stellar $\sigma_{z}$ as an input parameter, which would result in a higher value of stellar flaring in the outer disk.
Thus, we note that the increase in stellar disk thickness with radius is a robust result from our work.  

 Our detailed predictions for stellar flaring in the outer disk would provide a motivation to compare these results with future data, from GAIA, for example.
Interestingly, the various observational determination of the thickness with radius for the Milky Way outer disk so far do show flaring
(Momany et al. (2006) using 2MASS data, Lopez-Corredoira \& Molgo (2014) using SDSS-SEGUE data, Wang et al. (2018) using LAMOST data), which agree reasonably well with our results
for the net thickness (column 4, Table 2). However, the flaring values reported at larger radii by Lopez-Corredoira et al. (2014) are  higher  than our values.
This difference could arise if the effect of warp has not been correctly removed from the data, or because the data is contaminated by the thick disk stars (Minchev et al. 2015). Furthermore,  we point out that a  tidal encounter with an external galaxy can cause heating of the stellar disk, preferentially in the outer parts (Walker et al. 1996).
This could be one reason for the higher flaring seen in observed data.  
Finally, we note  that Wang et al. (2018)
 use a sech$^2$ profile to model the observed data to derive the scale height, while the real profile is different, as discussed above.

Thus our work has shown that the stellar disk is not rigorously constant, as has long been assumed in the literature (van der Kruit \& Searle 1981). 
Hence,  the subsequent relation  that $\sigma_{z}$ falls off exponentially with radius with a scale length of $2 R_D$ (where $R_D$ is the exponential disk scale length) as  proposed by them, which is routinely used in the literature, is not physically justified either.

We note that a moderate increase in stellar flaring in external galaxies has long been known from observations (de Grijs \& Peletier 1997). 
This trend was also seen  when the multi-component model was applied and the results from it compared quantitatively with the intensity profiles for NGC 891 and NGC 4565 (Narayan \& Jog 2002a).  However, this  has not received the attention it deserves. Our result for the flaring of the stellar disk in the outer Galaxy
agrees with the trend noted and discussed by Kalberla et al. (2014), who  have also questioned the constancy of the scale height with radius.

\subsubsection{z$_{1/2}$, the half-mass scale height}
We introduce a new parameter 
$z_{1/2}$, the half-mass scale height,  measured from the mid-plane, as another indicator of the constraining effect of gas and the halo. Here 2 $z_{1/2}$ contains half the total stellar surface density. This will also implicitly depend on the shape of the density profile at larger $z$ values, unlike the value of HWHM. Hence a smaller $z_{1/2}$ indicates a distribution that is concentrated closer to the mid-plane.
From an observational perspective, $z_{1/2}$ can be thought of as an  indicator of the half-light scale-height of an edge-on galaxy with $M/L$ considered to be constant.

The calculated values for $z_{1/2}$ from R= 4 kpc to 22 kpc are given in Table 3 as determined for the three cases considered.
This also shows the same trend,  namely, gas and  dark matter halo are more important at smaller and larger radii, respectively, and the values of
$z_{1/2}$ increase with radius. The actual values of $z_{1/2}$ are lower than the corresponding HWHM values  because their definitions are different. 
\begin{table}
\caption{$z_{1/2}$  of vertical stellar density distribution vs. radius
}
\label{table:3}
\centering
\begin{tabular}{l l l l}
\hline \hline
Radius & Stars-alone & Stars+Halo & Stars+Halo+Gas \\
(kpc) & (pc) & (pc) & (pc)\\ \hline
4.0  & 195.6  & 187.2  & 164.1\\
6.0  & 231.1  & 215.0  & 177.0\\
8.5  & 284.4  & 250.7  & 209.0\\
10.0 & 320.4  & 269.6  & 218.9\\
12.0 & 375.4  & 290.9  & 219.4\\
14.0 & 432.2  & 302.4  & 230.7\\
16.0 & 495.8  & 303.5  & 232.9\\
18.0 & 969.3  & 396.4  & 324.4\\
20.0 & 1544.6 & 480.7  & 416.6\\ 
22.0 & 1981.8 & 554.9  & 502.2\\ 
\hline
\end{tabular}
\end{table}
\subsection{Measuring the constraining effect: increased mid-plane density}

In a multi-component disk in the field of the halo, the mid-plane density is higher than in the one-component case (see Section 3.1).
The resulting mid-plane density values are given in Table 4. The increase in density is particularly important at larger radii where the dark matter halo force dominates  the force due to the stellar component itself (see Fig. 2). While the mid-plane density is not likely to be directly observed, the values give an idea of how the mid-plane stellar density  
increases in a coupled system.
The increase in the mid-plane density, which can be a factor of $\sim 4$ in the outer disk, can have interesting dynamical consequences, for example, 
the vertical oscillation frequency given as $(4 \pi G \rho_0)^{(1/2)}$ (see Binney \& Tremaine 1987) would be higher by a factor of $\sim 2$ and could thus lead to a better-mixed vertical distribution in the outer disk.

\begin{table}
\caption{Mid-plane density of stars vs. radius}
\label{table:1}
\centering
  \begin{tabular}{l l l l}
\hline \hline
Radius & Stars-alone & Stars+Halo & Stars+Halo+Gas \\
(kpc) & $(M_{\sun}\mathrm{pc^{-3}})$ & $(M_{\sun}\mathrm{pc^{-3}})$ & $(M_{\sun}\mathrm{pc^{-3}})$ \\
 \hline
4.0 & 0.258 & 0.269 & 0.313\\
6.0 & 0.117 & 0.125 & 0.156\\
8.5 & 0.043 & 0.049 & 0.06\\
10.0 & 0.024 & 0.029 & 0.035\\
12.0 & 0.011 & 0.014 & 0.019\\
14.0 & 0.005 & 0.007 & 0.01\\
16.0 & 0.002 & 0.004 & 0.005 \\
18.0 & 6$\times10^{-4}$ & 1.6$\times10^{-3}$ & 0.002 \\
20.0 & 2$\times10^{-4}$ & 7$\times10^{-4}$ & 8$\times10^{-4}$\\
22.0 & 9$\times10^{-5}$ & 3$\times10^{-4}$ & 4$\times10^{-4}$\\
\hline
\end{tabular}
   \end{table}

\subsection{Measuring the constraining effect: stellar vertical profile}

In real galaxies, the vertical stellar density profile close to the mid-plane does not follow the sech$^2$ profile expected of a one-component, isothermal disk (Spitzer 1942), but instead is close to a $sech$ distribution or lies between a $sech$ and an exponential distribution, while at large $z$ the profile tends to have an exponential form.
Van der Kruit (1988) suggested a family of curves that could fit these observational trends:
\begin{equation}
\rho\left(z\right)=2^{-2/n}\rho_{e}\sech^{2/n}\left(nz/2z_{e}\right).\end{equation}
Here $n=1,2$ and n$\rightarrow \infty$ correspond to a density profile with sech$^2$, sech and an exponential $z$ 
distribution, respectively.

The parameters $n, z_e$ , and $\rho_e$ are obtained  with a best-fit analysis of the data.
 We note that this profile is ad hoc and is not based on any physical principle, as was noted by van der Kruit (1988). Many
observational studies have presented their results in terms of this function and have given the parameter $n$ for the best-fit case (e.g., Barteldrees \& Dettmar 1994; de Grijs et al 1997).

\begin{figure}
\centering
\includegraphics[height=2.0in,width=2.8in]{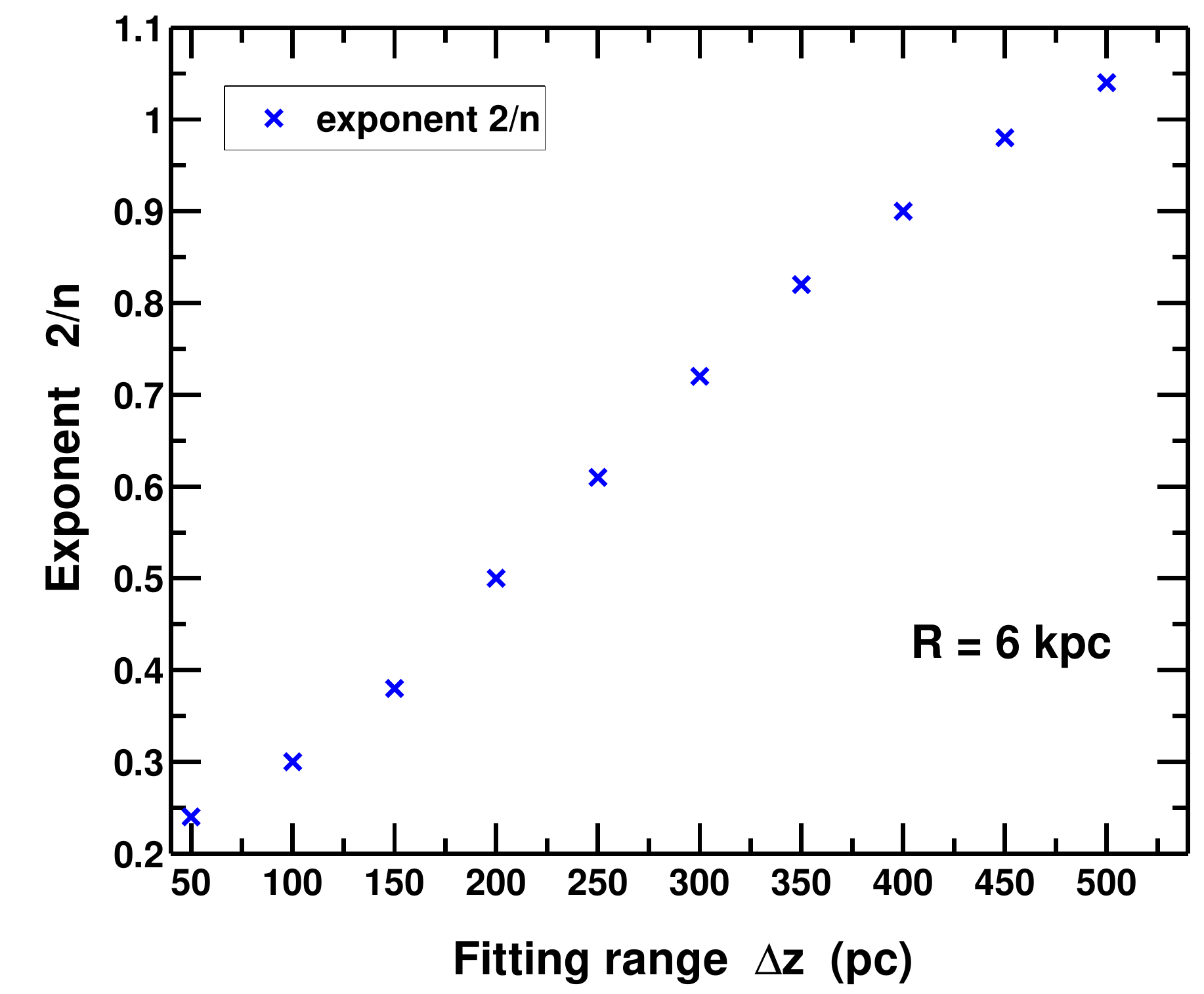}
\caption{Variation in best-fit value of the exponent $2/n$ as a function of fitting range $ \Delta z$, when our results for stellar 
density distribution at R=6 kpc are fit with a distribution of type sech$^{2/n}$.  The value of the exponent $2/n$  varies with $\Delta z$,  hence it cannot be regarded as a robust indicator of stellar density profile.}
\label{label5}
\end{figure}

We next try to fit the resulting stellar density profile from our work (Section 3.1)
 by the above function (Eq. (8) ).  We note that $2^{-2/n} \rho_e$ is the mid-plane stellar density, which is not known in observational studies, but we know it from our model calculations.
 For each radius, we obtain the best-fit value for $n$ and $z_e$. Interestingly, we find that the best-fit value of the exponent $2/n$ is not robust, but instead depends on the z-range ($\Delta z$) chosen for the fitting, as shown in Fig. 5 for the case of R=6 kpc. For example, the value of $n$ thus obtained is  $\sim 5$ when $\Delta z$= 150 pc, while $n \sim 2 $ for $\Delta z$= 450 pc. These would physically imply
very different density profiles of $~sech^{0.4}$ and $~sech$ respectively (see Eq. (8)).
This clearly shows that the family of curves suggested by van der Kruit (1988) for the density profile is not valid physically for a realistic disk treated here. 
 Hence for
a working definition of disk thickness,  we have used the HWHM value here, regardless of the actual shape of the density profile.

However, keeping in mind  this caveat that $n$ is not robust, we next try  to use this function to fit our resulting stellar distribution vs. $z$ (Section 3.1)  within $|z| < 150$ pc so as to estimate the trend seen, and to be able to compare this with observational studies where data close to the mid-plane are studied. 
 The resulting plot of $2/n$ vs. radius is shown in Fig. 6.

\begin{figure}
\centering
\includegraphics[height=2.0in,width=2.8in]{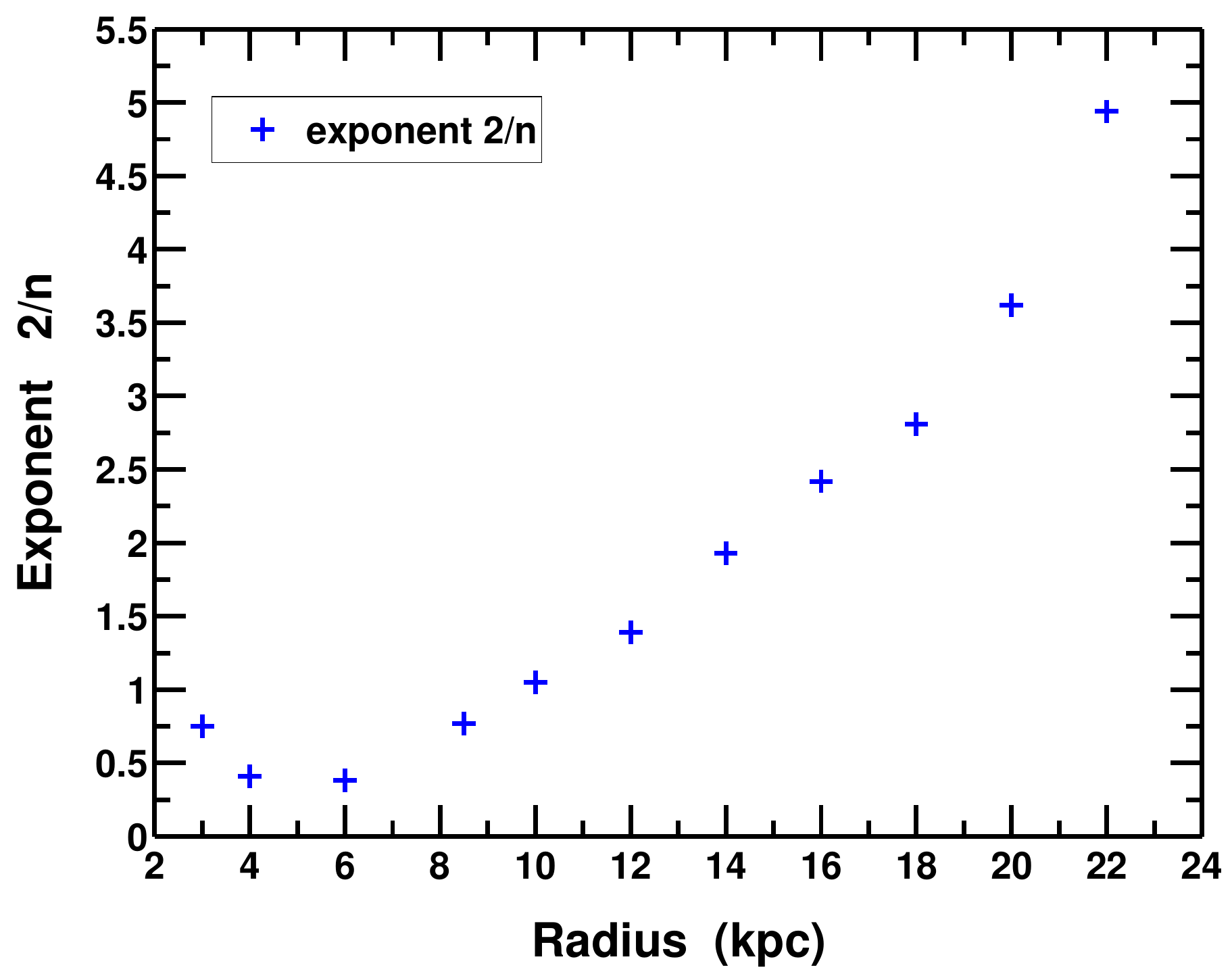}
\caption{Best-fit value of the exponent {2/n} vs. radius when our results for the stellar density distribution for the Milky Way are fit by a distribution of type sech$^{2/n}$ at each radius. For radii $>$ 14 kpc,  a new regime of $n < 1$ is obtained. This corresponds to radii where the main constraining  effect is due to the  dark matter halo.}
\label{label6}
\end{figure}

Interestingly, Fig. 6 shows that our results for the stellar density profiles gives $n > 1$ in the region $R < 14 $ kpc 
when the constraining of the stellar profile is mainly due to gas (Section 3.2), whereas we find $ n < 1$ in the outer Galactic disk when the constraining is due to the dark matter halo. The typical observed profiles for external galaxies are for the inner region of the disk where the constraining effect of gas dominates, and hence our calculations explain why the observed values measure $n$ to be $> 1$ with density profiles between $sech$ and an exponential.
When the additional gravitating component is more massive and extended along $z$ than the stars  (such as the halo), it can affect the stellar profile over a larger $z$ range, and this results in $n < 1$.   
This regime of $n < 1$ has so far not been studied in the literature.  An interesting point to note is that even if the stars are taken to be isothermal, the net stellar distribution is no longer given by a $sech^2$ profile in a multi-component disk and under the force of the dark matter halo.

In the literature the value $n > 1$ or correspondingly $2/n < 2$ is often taken to be an indicator of the steepening of the stellar density profile (e.g., de Grijs et al. 1997; Banerjee \& Jog 2007). We note that this labeling is misleading, since the case where $n< 1$ could 
also lead to a steeper profile.
  In reality, all the three parameters (the mid-plane density, width and $n$ ) decide how sharply the density profile falls with $z$.
As we have shown in Section 3.1, for a given constant surface density,  an additional component will always cause a steepening of the stellar profile. Similarly, the other disk components (HI and H$_2$ gas) will also
show a steeper profile than the corresponding one-component cases.

\section{Vertical distribution of HI}

\begin{figure*}
\centering
\includegraphics[height=2.0in,width=2.8in]{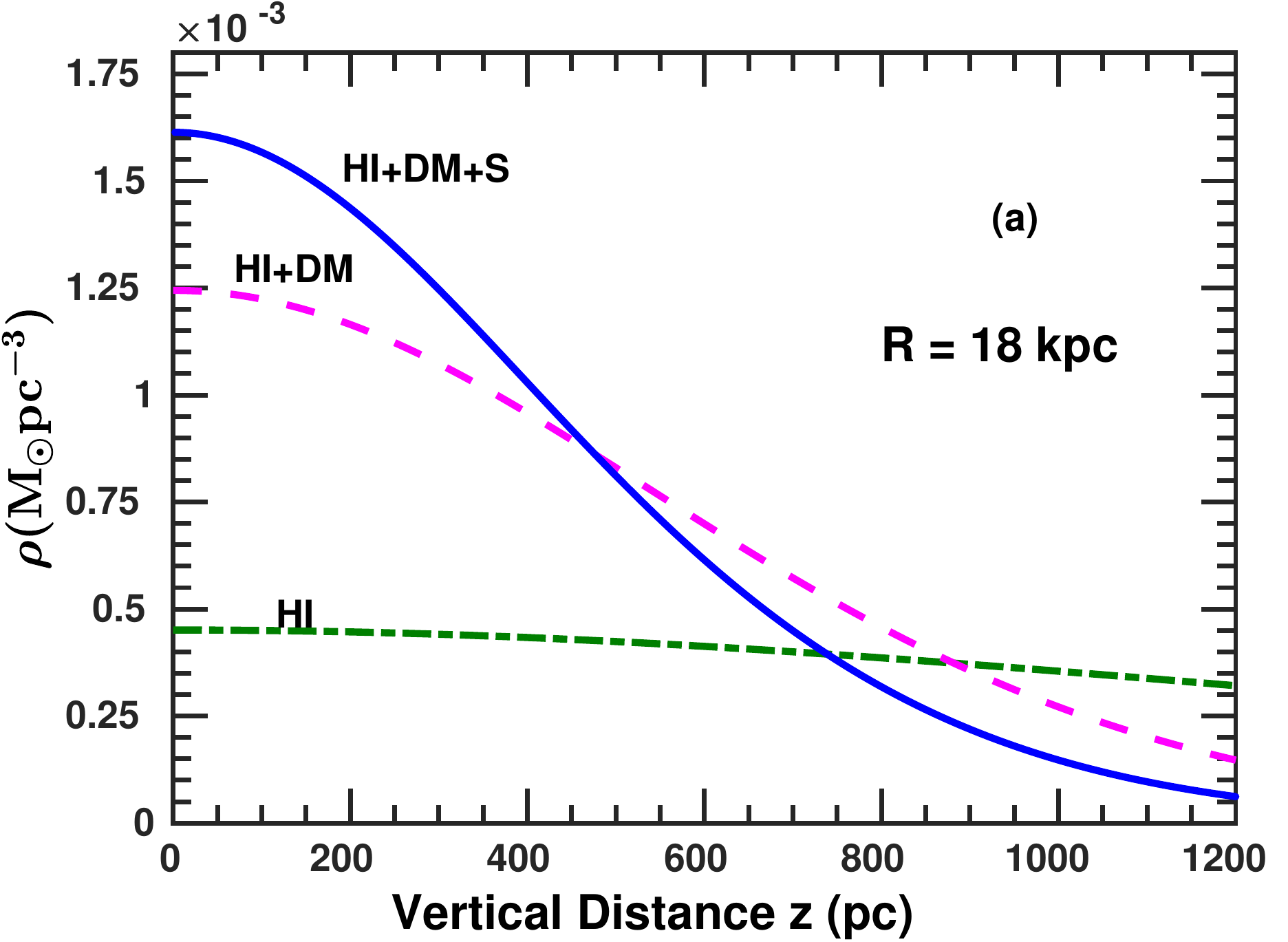}
\medskip
\includegraphics[height=2.0in,width=2.6in]{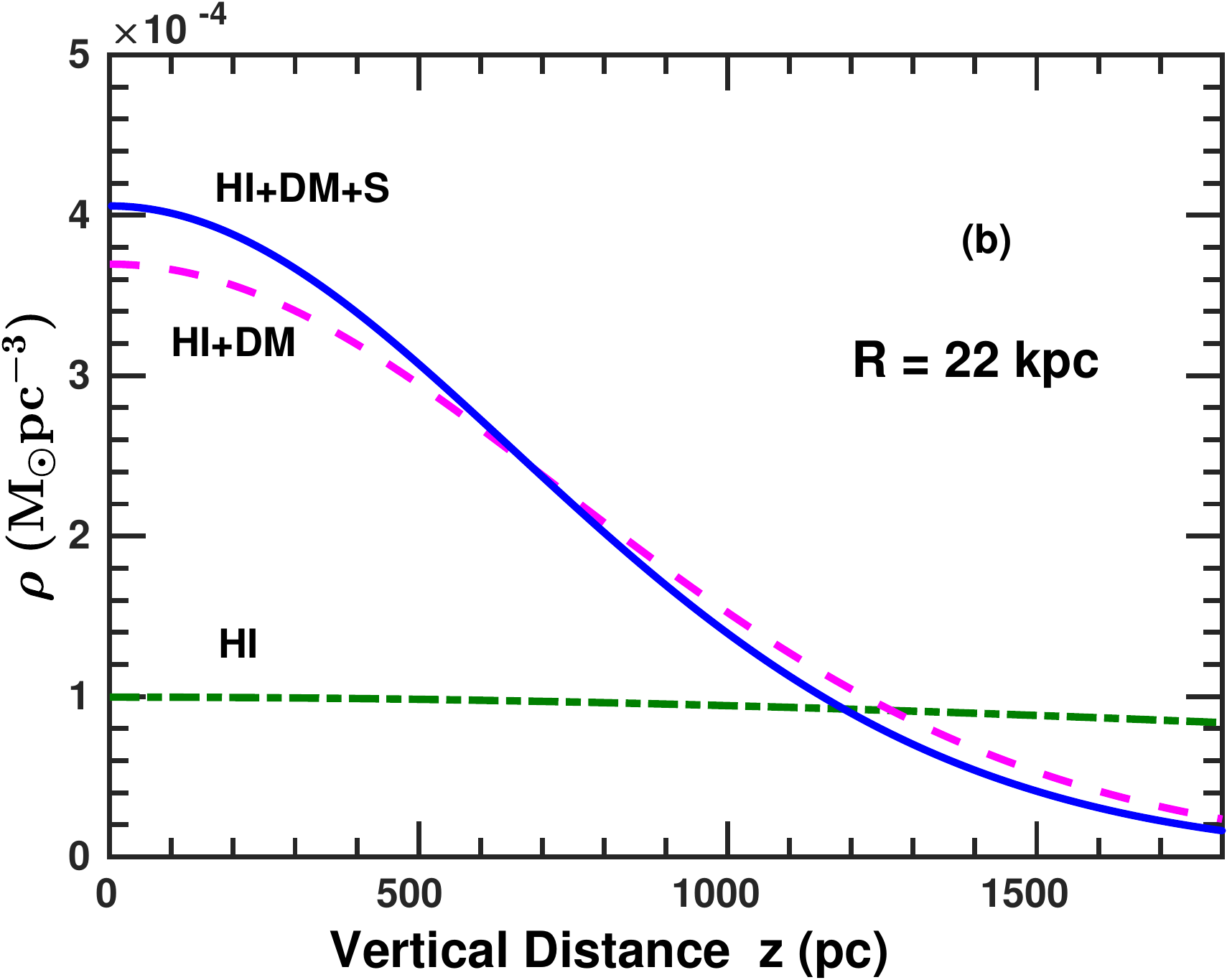}
\bigskip
\caption{Vertical density distribution of HI gas under its own self-gravity, then also including the gravitational force of the dark matter halo, and then including the effect of both halo and stars, at R= 18 kpc (a) and at R= 22 kpc (b). The halo very strongly constrains the HI distribution in the outer Galaxy, and the effect is progressively higher at  larger radii.}
\end{figure*}

Although the focus of this paper is on the resulting stellar density distribution, in this section we briefly consider  the resulting HI gas density distribution in the outer disk for the sake of completeness. The HI scale heights (defined as the HWHM of the HI vertical density distribution) for the region inside of R= 12 were considered in detail by Narayan \& Jog (2002b), hence we do not consider that range here.

The procedure for obtaining the stellar distribution (by solving the coupled Eq. (1)) also simultaneously gives the gas vertical distribution. We also consider the HI gas plus halo case. For comparison, we obtain the HI-alone distribution analytically (Section 3.1). 
The corresponding plots for R=18 kpc and R=22 kpc are given in Fig. 7. 
The most striking result form this figure is that the gas-alone case under its own gravity would be highly extended vertically with an HWHM of 
1.8 kpc and 3.8 kpc at R=18 kpc and 22 kpc, respectively. These are much higher than the observed values (see, e.g., Wouterloot et al. 1990).
Including the gravitational field of the dark matter halo substantially reduces
 the HMHM values,  by a factor of 2.7 and 4.3 at these two radii, respectively. The decrease is slightly stronger when the effect of stars is also included, so that the net HWHM values are 502 pc and 798 pc, respectively. Interestingly, these resulting values of the HWHM (in a multi-component disk plus halo) up to R=22 kpc agree fairly well with the observed values of gas flaring (Wouterloot et al. 1990,  Kalberla \& Dedes 2008), while Levine et al. (2006) give higher values. 
The HI gas thickness shows an azimuthal variation (Kalberla \& Dedes 2008, Levine et al. 2006), which may be explained in terms of  intergalactic accretion flows, as proposed by 
Lopez-Corredoira \& Betancort-Rijo (2009).

The corresponding mid-plane density for HI increases by a factor of 3.6 and 4.1, respectively, compared to the gas-alone values,
and the corresponding net mid-plane density values are 1.6$\times10^{-3}$ and 4.1$\times10^{-4} M_{\odot} \mathrm{pc^{-3}}$.
Thus without the confining effect of halo, the vertical distribution of HI would be very extended, which would then be even more susceptible to  disturbance  by perturbations, including gas dynamical processes, than stars.

While the net gravitational force due to the halo decreases with radius, the gas velocity dispersion $\sigma_{z}$ is nearly constant and hence the gas
scale height shows huge flaring in the outer Galaxy (see Fig. 7). This is despite the strong constraining effect of the halo.
This flaring is avoided in the inner Galaxy by the strong gravitational field of stars and gas,
which results in a near-constant HI scale height in the inner Galaxy, in agreement with observations, as was shown by Narayan \& Jog (2002b).

\section{Conclusion}
 In summary, we have studied the vertical stellar distribution of the Milky Way thin disk in the outer disk (R=12-22 kpc) by applying the method of Narayan \& Jog (2002b) for a multi-component disk in the field of the dark matter halo. We have shown that the dark matter halo plays a dominant role in reducing the disk thickness in the outer Galaxy.

\medskip

\noindent The main results are summarized below:

\noindent 1. The thickness of the stellar disk (measured as the HWHM of the vertical density distribution) increases moderately by $\sim 40  \% $ up to R= 12 kpc, and flares at larger radii up to $\sim 1$ kpc at R=22 kpc. The stars-alone disk would flare
even more. Including the gravitational effect of the halo restricts the stellar distribution to be closer to the mid-plane at 
 large radii, which may help the disk resist distortion due to external perturbations.
The overall trend of flaring of the stellar disk in the outer Milky Way is a robust result from our work.

\noindent 2. The mid-plane stellar density is higher in a multi-component disk in the field of the halo than for the stars-alone case.
The increase is a factor of $\sim 4$ in the outer disk.

\noindent 3. Because of   the gravitational effect of gas and  dark matter halo, the stellar density distribution 
falls more steeply than for a stars-alone disk.
We showed that close to the mid-plane, the stellar density profile varies as sech$^{(2/n)}$, where $ n < 1$ when constrained by the dark matter halo. 
The regime of $ n < 1$  has so far not been studied in the literature. 

\noindent 4. The HI gas in the outer Galaxy also shows flaring, and without the dark matter halo, the flaring would be even larger. Such
extended gas at higher $z$ values would be liable to  get disturbed more easily by perturbations. 

The formulation in this paper is general. Hence we expect the trends in results, namely the flaring of the disk, and the dominant role played by the dark matter halo in constraining the outer disk, to be applicable for other galaxies as well.

Thus, the dark matter halo that is normally evoked to explain the nearly flat rotation curves in galaxies also strongly affects
the stellar and gas distribution normal to the disk in the outer parts of the Milky Way, causing them to be confined closer to the disk plane. 
This dynamical effect of the halo on the disk distribution is
progressively more important at larger radii. 
 Thus it is crucially important to include the 
effect of the dark matter halo while studying the vertical structure and dynamics of a galactic disk. 

\medskip

\noindent {\bf Acknowledgments:}  We would like to thank the anonymous referee for constructive comments that have helped
us improve the paper. SS would like to thank CSIR for a fellowship. CJ would like to thank the DST, Government
of India for support via a J.C. Bose fellowship (SB/S2/JCB-31/2014).

\bigskip

\noindent {\bf {References}}
\medskip

\noindent Banerjee,A., \& Jog, C.J. 2007, ApJ, 662, 335

\noindent Banerjee, A. \& Jog, C.J. 2013, MNRAS, 431, 582

\noindent Bartledrees, A. \& Dettmar, R.-J. 1994, A\&AS, 103, 475

\noindent Binney, J. \& Merrifield, M. 1998, Galactic Astronomy (Princeton: Princeton Univ. Press)

\noindent Binney, J., \& Tremaine, S. 1987, Galactic Dynamics
(Princeton:Princeton Univ.Press)

\noindent Comeron, S., Elmegreen, B.G., Knapen, J.H et al. 2011, ApJ, 741, 28

\noindent de Grijs, R., \& Peletier, R.F. 1997, A\&A, 320, L21

\noindent de Grijs, R., Peletier, R.F., van der Kruit, P.C. 1997, A \& A , 327, 966

\noindent Dehnen, W., \& Binney, J. 1998, MNRAS, 298, 387

\noindent Dickey, J.M. 1996, in Unsolved problems of the Milky Way, ed. L. Blitz and P. Teuben
   (Dordrecht: Kluwer), IAU Symp. 169, 489

\noindent Jog, C.J., \& Narayan, C. A. 2001, MNRAS, 327, 1021

\noindent Kalberla, P.M.W., \& Dedes, L.,  2008, A\&A, 487, 951

\noindent Kalberla, P.M.W.,  Kerp, J.,  Dedes, L., \&   Haud, U. 2014, ApJ, 794, 90

\noindent Kamphuis, J.J. 1993, Ph.D. thesis, University of Groningen

\noindent Levine,E.S., Blitz,L., \& Heiles,C. 2006, ApJ, 643, 881

\noindent Lewis, B.M. 1984, 1984, ApJ, 285, L453
 
\noindent Lewis, J.R., \& Freeman, K.C. 1989, AJ, 97, 139

\noindent Lopez-Corredoira, M, \& Betancort-Rijo, J.  2009, A\&A, 493, L9

\noindent Lopez-Corredoira, M, \& Molgo, J. 2014, A\&A, 567, A106
        
\noindent Mera, D., Chabrier, G., \& Schaeffer, R. 1998, A\&A, 330, 953
 
\noindent Minchev, I., Martig, M., Streich, D. et al.  2015, ApJ, 804, L9

\noindent Mignard, F. 2000, A\&A, 354, 522

\noindent Momany, Y., Zaggia. S., Gilmore. G. et al.  2006, A\&A, 451, 515

\noindent Narayan, C.A., \& Jog, C.J. 2002a, A\&A, 390, L35

\noindent Narayan,C.A., \& Jog, C.J. 2002b, A\&A, 394, 89

\noindent Narayan,C.A, Saha, K., \& Jog,C.J. 2005, A\&A, 440, 523

\noindent Scoville, N.Z., \& Sanders, D.B. 1987, in Interstellar 
Processes, eds. D.J. Hollenbach \& H.A. Thronson (Dordrecht: Riedel), 21

\noindent Spitzer, L. 1942, ApJ, 95, 329

\noindent Spitzer, L. 1978, Physical Processes in the Interstellar Medium
 (New York:John Wiley)

\noindent Tamburro, D., Rix, H.-W., Leroy, A.K. et al.  2009, AJ, 137, 4424

\noindent van der Kruit, P. C., \& Searle, L. 1981, A\&A, 95, 105

\noindent van der Kruit, P. C. 1988, A\&A, 192, 117

\noindent Walker, I.R., Mihos, J.C., Hernquist, L. 1996, ApJ, 460, 121 

\noindent Wang, H.-F., Liu, C., Xu, Y., Wan, J.-C., Deng, L. 2018, ArXiv 1804.10485

\noindent Wouterloot, J.G.A., Brand, J., Burton, W.B., Kwee, K.K. 1990, A\&A, 230, 21

\noindent Young, J.S., Scoville, N.S. 1991, ARAA, 29, 581

\end{document}